**Chemical Vapor Deposition of Epitaxial Chromium Nitride Thin Films**

**L. J. Adams[1*], S. Baserga[1], L. Souqui[1], E. Sadek[1], L. von Fieandt[1,2], P. Eklund[1]**

[1]Department of Chemistry – Ångström Laboratory, Uppsala University, SE-75120 Uppsala, Sweden.

[2]AB Sandvik Coromant, SE-12679 Hägersten, Sweden.

**\*Corresponding author: lewis.adams@kemi.uu.se

**Abstract**

Chromium nitride (CrN) is a thermoelectric transition metal nitride whose properties are strongly influenced by stoichiometry, substrate choice, and defect chemistry. CrN is routinely synthesized by physical vapor deposition (PVD), its growth by chemical vapor deposition (CVD) has been limited by the lack of suitable chromium precursors capable of producing carbon-, oxygen-, and chlorine-free films. CVD of contamination-free Cr compounds is notoriously difficult, with carbon-free Cr compounds thought unattainable below 1000 °C. Here, we report epitaxial, carbon- and chlorine-free CrN thin films synthesized by thermal CVD. Single-phase CrN films (~110 nm) were deposited on c-plane $\alpha$-$Al_2O_3$ using *in-situ* generated chromium chlorides. Films exhibit n-type conduction with a Seebeck coefficient of −36 $\mu V\ K^{-1}$, comparable to PVD-grown CrN. These results present a route to highly crystalline rock-salt CrN films for defect engineering, doping, and alloying with reduced implantation-related damage capabilities previously largely confined to PVD-based techniques.

**Table of contents Image**

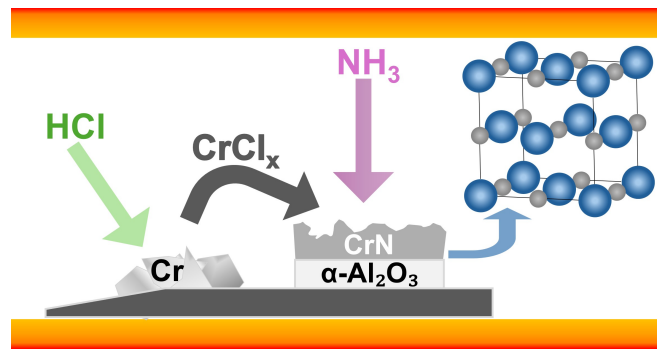

**Main Article**

Rock-salt chromium nitride (CrN) is long-established as a hard[1–3], wear-resistant[4,5], and corrosion-resistant coating. In its bulk form, it is known as Carlsbergite and occurs naturally in meteorites.[6] In recent years, thin films of CrN have also shown promise as thermoelectric materials.[7,8] In contrast to conventional thermoelectrics such a $Bi_2Te_3$ or PbTe, CrN is based on abundant raw materials.[9,10]

CrN thin films have long been synthesized using physical vapor deposition (PVD) techniques such as cathodic arc deposition[11–13] and magnetron sputtering.[14–16] In contrast, chemical vapor deposition (CVD) of CrN remains essentially unexplored, with only a few studies reporting the use of organometallic precursors.[17,18] The reason why such a seemingly straightforward material is challenging to synthesize by CVD is due to the lack of suitable chromium precursors that produce carbon-, oxygen-, or chlorine-free films. In fact, carbon-free deposition of Cr by CVD at low temperatures is described as essentially "impossible," with some textbooks stating that such deposition cannot be achieved below 1000 °C.[19,20] For the few examples of CVD of CrN there are, using organometallics results in coatings suffering from high levels of carbon- and oxygen-contamination.[17,18] Alternatively, chromium-halide-based-precursors, particularly chromium chlorides, have not been explored for CVD of CrN due to their high volatilization temperatures (>500 °C)[21,22], which hinder efficient precursor delivery.

*In-situ* precursor generation has been used for the deposition of metallic Cr, employing hydrogen chloride (HCl) as chlorinating agent[23,24], but this strategy has not previously been applied to CrN. Thermal CVD is a technique known for its excellent conformality and step coverage,[25] making it ideal for coating complex substrates and devices. In this study, we report the implementation of *in-situ* generated chromium chlorides, formed through the reaction of metallic chromium with HCl, as an effective chromium precursor for the deposition of single-phase epitaxial CrN thin films. The structure, composition, and epitaxy of the as-deposited films were investigated using High-Angle Annular Dark Field Scanning Transmission Electron Microscopy (HAADF-STEM), Time-of-Flight Elastic Recoil Detection Analysis (ToF-ERDA) and X-Ray Diffraction (XRD).

Films were deposited using an in-house-built three-zone horizontal hot-wall tube reactor. A detailed decription of the reactor is available in the thesis of Gerdin Hulkko.[26] The reagents (Ar (99.9999%), $H_2$ (99.9996%), HCl (99.0%), $NH_3$ (99.999%)) were delivered through Inconel 600 injectors and passed through heating zone 1 entering the reactor, see Figure 1. HCl and $NH_3$ were used as a chorinating agent and nitrogen source, respectively. A TiN-coated graphite sample holder containing cleaned, single-side polished c-plane sapphire substrates (see details in the Supporting Information) and 2.5 g of metallic chromium (~4-6 mm in lateral size) was placed in heating zone 3, furthest away from the injectors. Heating zones 2 and 3 were preheated to 700 °C, while heating zone 1 was kept at the lowest feasible temperature of approximately 350 °C in order to prevent undesired etching of the precursor injectors. The chamber pressure was maintained at 213 Pa (1.6 Torr), with a total gas flow of 700 SCCM, yielding a linear flow velocity of 19 m $s^{-1}$. Individual gas flows of 75 SCCM and 50 SCCM corresponded to an $NH_3$:HCl ratio of 3:2; an excess of ammonia was used to reduce the likelihood of $Cr_2N$ metallic phase formation. 200 SCCM of $H_2$ were used to facilitate the reduction of the chromium-generated chloride species. The individual gas flows are shown in the supporting information (Table S1).  After deposition, the reactor was cooled for 30 minutes under vacuum while maintaining an argon flow of 400 SCCM. Following this period, the argon flow was stopped, and the reactor was allowed to cool gradually to room temperature under vacuum alone.

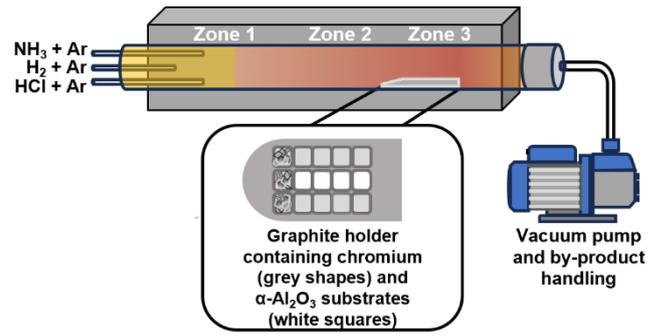

**Figure 1:** Schematic diagram of the in-house-built hot-wall CVD tube reactor.

Figure 2(a) shows a $\theta$-$2\theta$ XRD pattern of CrN deposited on c-plane sapphire. The pattern shows peaks at 37.52° and 80.07°, which correspond to the 111 and 222 reflections, respectively, of the rock-salt cubic CrN structure. No other peaks, apart from those from the substrate, were observed. The film exhibits a strong preferred out-of-plane orientation along the [111] direction. The rocking curve (Figure S1) of the 111 reflection yielded a FWHM of 0.1842°, indicating a low out-of-plane mosaic spread, which is consistent with high crystalline quality.

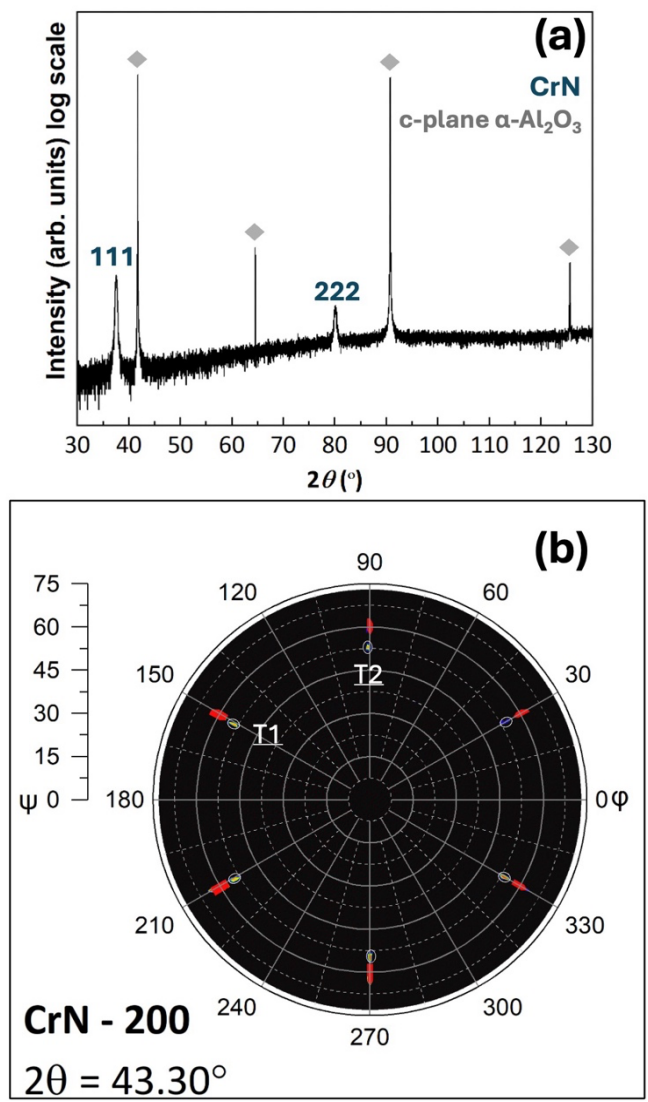

**Figure 2: (a)** θ-2θ X-ray diffraction pattern of CrN on c-plane sapphire. **(b)** Pole figure of 200 CrN poles measured at 2θ = 43.30°. The two in-plane twin variants observed in the CrN film are labelled T1 and T2. The pole figure is plotted on a log intensity scale. The poles corresponding to CrN are highlighted with white circles.

Figure 2(b) shows an XRD pole figure of the 200 CrN reflection, acquired at 2θ=43.30°. In Figure 2(b), reflections belonging to the sapphire $\{11\bar{2}3\}$ family appear at Ψ=60° every 60° of rotation around ϕ which is to be expected due to their sixfold symmetry. Reflections belonging to the {200} family of planes for CrN appear at Ψ=54° every 60° of rotation along ϕ. These planes have a threefold symmetry; thus, the poles are expected to be observed every 120° of rotation around ϕ. The fact that six peaks are observed shows twinning of the CrN, as often observed for NaCl-type of structures on c-plane sapphire.[27–29] Thus, the pole figures confirm the epitaxial growth of CrN on c-plane sapphire with in-plane twinning; the epitaxial relations are CrN[$11\bar{2}$] // α-Al$_2$O$_3$[$10\bar{1}0$] and CrN[$\bar{1}\bar{1}2$] // α-Al$_2$O$_3$[$10\bar{1}0$] in-plane. Additionally, pole figures of the 111 CrN reflection, 104 α-Al$_2$O$_3$ reflection and the 101 Cr$_2$N reflections are provided separately in the Supporting Information. No poles were observed for the Cr$_2$N 101 reflection, further confirming the absence of the nitrogen deficient phase.

Figure 3 shows STEM images and EDS compositional maps of the CrN thin films. The films are entirely composed of chromium and nitrogen, with only trace amounts of oxygen detected in the film. From the STEM image, it is evident that the CrN grains did not fully coalesce during the 4 hour deposition period, with the film measuring a thickness of approximately 110 nm. Dark regions within the film can be observed near the base of the films potentially caused by defects near the substrate-film interface or as the result of STEM edge effects due to poor wetting of the Pt capping layer. No trace metals (*e.g.,* Ni, Fe), chlorine, or carbon contamination were detected, indicating that HCl-induced etching of the gas injectors did not occur or affect the composition of the film, and that *unreacted* chromium chlorides were not incorporated into the film.

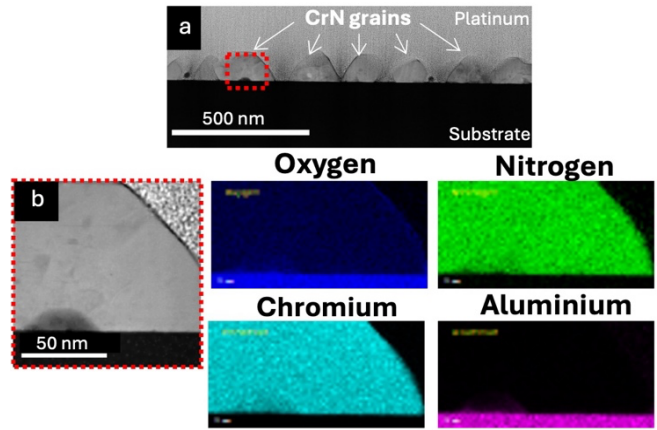

**Figure 3: (a)** Scanning transmission electron microscopy (STEM) image of a CrN film deposited at 700 °C. **(b)** Corresponding energy-dispersive X-ray spectroscopy (EDS) elemental map of the same region. The magnification scale in **(b)** is identical for all EDS images shown.

The film composition was quantified by ToF-ERDA, yielding an average Cr:N ratio of 1.00:0.91, indicative of the presence of nitrogen vacancies ($V_N$). This nitrogen deficiency is attributed to limited $NH_3$ cracking efficiency and kinetically constrained nitrogen incorporation, despite $NH_3$ rich gas-phase conditions. Under such slightly nitrogen-rich environments, chromium vacancies would typically be expected due to their lower formation energy, as reported by le Febvrier *et al*.[30]  Additionally, at the elevated deposition temperature, nitrogen may desorb from the growing film as $N_2$, favoring $V_N$ formation while the film retains the rock-salt CrN structure owing to its low nitrogen vacancy formation energy and high tolerance to point defects.[31] No traces of chlorine were detected in the films from the ToF-ERDA, further evidence that all the chromium chlorides generated *in-situ* were fully consumed near the surface of the substrate and not incorporated into the film, further verifying the high quality of the nitride. It was estimated that the CrN contained ~4 at.% of oxygen which is comparable to values observed in high-vacuum PVD-produced CrN films (~1–4 at.%)[15,32] and consistent with low-pressure CVD rock-salt structure nitride films, e.g., tantalum nitride (~5 at.%) and titanium nitride (~3.5–10 at.%).[33,34] The presence of oxygen may originate from post-deposition passive oxidation[35], the low-pressure growth environment, or possible enrichment from the substrate surface induced by elevated temperature and HCl gas flow.[36,37] However, the oxygen concentration measured by ToF-ERDA is likely an overestimation of the true bulk oxygen content. This is due to the not fully coalesced nature of the film, resulting in the ion beam measuring contributions from both the film and the substrate.

The room-temperature resistivity as measured by four-point probe of the CrN films was 119 ± 30 mΩ·cm, which falls in the range of than values (1-300 mΩ·cm) reported for CrN films.[15,30] This value is slightly lower than pure stoichiometric rock-salt CrN. This reduction is attributed to the high concentration of nitrogen vacancies in the present films (Cr:N ≈ 1.0:0.9), which act as donor defects and increase the carrier concentration. The resulting band-gap narrowing and partial closure drive the material toward more metallic characteristics.[38] Similar behavior has been reported for nitrogen deficient CrN films grown by reactive magnetron sputtering.[30] Furthermore, we do not observe any $Cr_2N$ phase, a metallic phase that has much lower resistivity, of the order of ~0.05–0.83 m$\Omega$·cm.[39] For our films, a Seebeck coefficient ($\alpha$) of - 36 $\mu$V·K$^{-1}$ was measured under ambient conditions, which is comparable with previously reported sputtered CrN.[14,28] The negative value of the Seebeck coefficient indicates that the material exhibits n-type conduction, providing further evidence for the presence of nitrogen vacancies acting as donor defects. Further optimisation of the film composition and growth conditions is required to enhance the electrical conductivity for thermoelectric applications, thereby maximizing the power factor and dimensionless figure of merit. This study highlights an emerging avenue for the functionalisation of CrN films.

In summary, this work demonstrates the epitaxial CVD growth of single phase, rock-salt, n-type CrN on c-plane sapphire, with no measurable chlorine or carbon contamination typically associated with chromium-based films deposited by CVD. This is an exciting result, as it opens new pathways for defect tuning, doping, and alloying of CrN films under near-equilibrium growth conditions with fewer implantation-related defects. To date, such control has largely been achievable only through PVD-based techniques.

**Experimnetal Methods**

**Substrate Preparation**

Films were deposited on cleaned, single-sided polished c-plane $\alpha$-Al$_2$O$_3$ substrates (10 mm × 10 mm × 0.5 mm, Alineason). The substrates were cleaned in an ultrasonic bath using acetone and ethanol for 5 minutes in each solution, followed by blow-drying with nitrogen gas.

**STEM Anlaysis**

A transmission electron microscopy (TEM) thin foil specimen of the area of interest was produced using the focused ion beam (FIB) *in-situ* lift out technique described by R. M. Langford and C. Clinton[40]. A Helios Nanolab 650 scanning electron microscope using a Ga$^+$ ion source was used for the sample preparation.

The TEM specimens were analyzed using a CS- and Probe- corrected FEI Titan transmission electron microscope equipped with a Schottky field emission gun (FEG) operated at 300 kV. The microscope was operated in scanning mode (STEM) and a high angle annular darkfield detector was used to collect images. A Bruker Super X energy dispersive X-ray (EDS) detector was used for collection of EDS data.

**X-ray diffraction, pole figure and rocking curve measurements**

The degree of orientation for both coatings was measured by X-ray diffraction (XRD) rocking curve measurements ($\omega$-scans). A Philips MRD X-pert diffractometer equipped with a primary hybrid monochromator and diffracted beam mirror was used for the measurements. The instrumental resolution was determined to be 0.0159°, based on the FWHM of the sapphire 006 substrate reflection. In addition, pole figures in the $\Psi$ range -1–73° were acquired using a Philips MRD-XPERT diffractometer operating in point focus mode and equipped with a primary X-ray poly-capillary lens with crossed slits, a secondary flat graphite monochromator with parallel plate collimator, and a nickel filter. Cu K$\alpha$-radiation was used for all X-ray measurements. Pole figure measurements were performed on the CrN 200 reflection(2θ = 43.30°) and the CrN 111 reflection (2θ = 37.55°), as well as on the $\alpha$-Al$_2$O$_3$ 104 substrate reflection (2θ = 35.15°). In addition, measurements were carried out at the Cr$_2$N 101 reflection (2θ = 29.31°).

**ToF-ERDA:**

The coating composition was analyzed using time-of-flight elastic recoil detection analysis (ToF-ERDA), carried out with a 5-MV 15-SDH-2 Pelletron accelerator at the Tandem Laboratory, Uppsala University.(41) A 36 MeV iodine ion beam ($^{127}I^{8+}$) was employed at an incidence angle of 22.5° relative to the sample surface. Recoiled species were detected at a scattering angle of 45° using a time-of-flight detector coupled to an energy analyzer. Elemental depth profiles were evaluated using the *Potku* software, from which the chemical composition was determined.(42)

**Seebeck coefficient measurement:**

The room temperature Seebeck coefficient (α) was measured by a custom built thermoelectric setup previously described by Xin *et al.*. (43) The Seebeck coefficient $\alpha$ defined as $\frac{\Delta V}{\Delta T}$ was calculated from a temperature range of $T_H = 34.1\ ^oC$ to $T_C = 24.9\ ^oC$.

**4-point probe resitivity measuments:**

Resitivity was measured using an Ossila 4-point probe station under ambient conditions with no illumination.

**Supporting Information**

Experimental gas flow parameters for CrN deposition, together with rocking curve measurements of single-phase CrN and the α-$Al_2O_3$ substrate. Pole figures corresponding to the 111 CrN, 104 α-$Al_2O_3$ and 101 $Cr_2N$ poles.

**Table S1:** Experimental gas flows and purities for the CrN deposition.

| Gas (Purity) | Flow (SCCM) |
|---|---|
| Ar (99.9999%) | 150 |
| | 125 |
| | 100 |
| $H_2$ (99.9996%) | 200 |
| HCl (99.0%) | 50 |
| $NH_3$ (99.999%) | 75 |

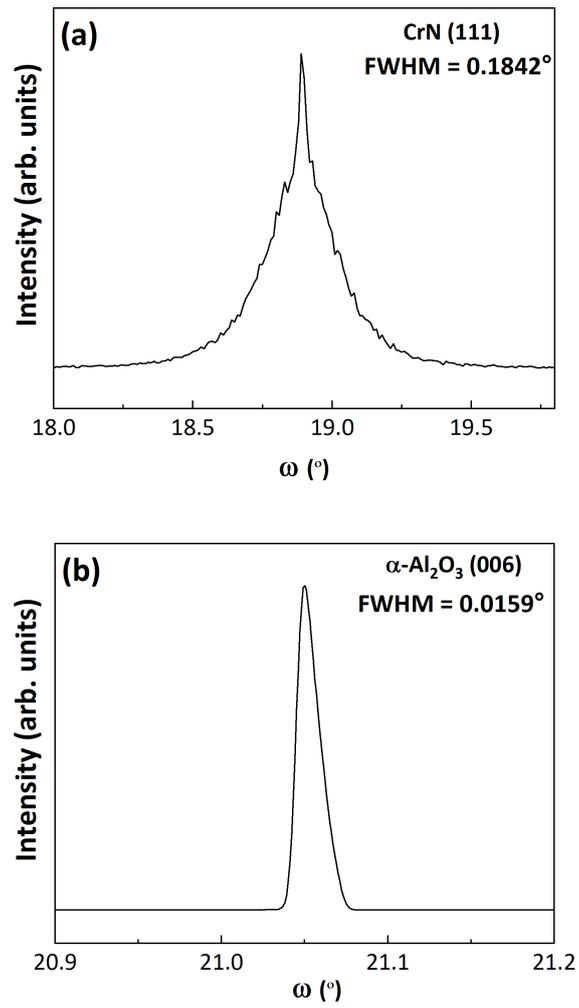

**Figure S1**: Rocking curve measurements of single phase CrN deposited at 700 °C. (a) Rocking curve scan of CrN (111). (b) Rocking curve scan of c-plane α-Al$_2$O$_3$ (006).

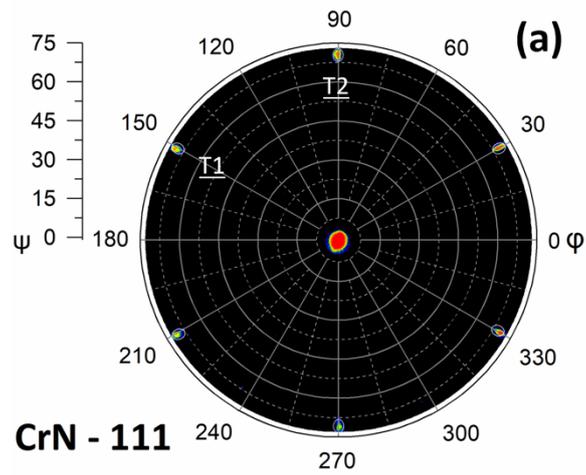

**CrN - 111**
2θ = 37.55°

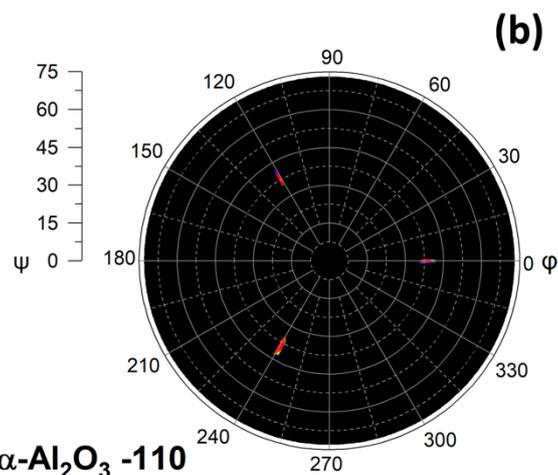

**α-Al₂O₃ -110**
2θ = 35.15°

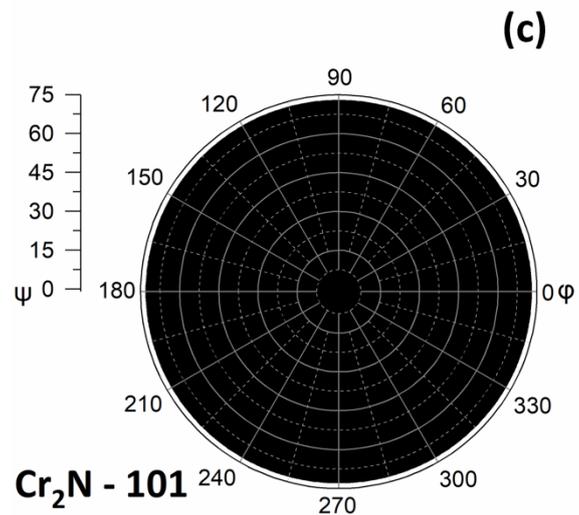

**Cr₂N - 101**
2θ = 29.31°

**Figure S2: (a)** Pole figure of 111 CrN pole measured at 2θ = 37.55°. The two in-plane twin variants observed in the CrN film are labelled T1 and T2. The poles corresponding to CrN at 70.5° highlighted with white circles. **(b)** Pole figure of c-plane α-Al₂O₃ measured at 2θ = 35.15°. **(c)** Pole figure measured at 2θ = 29.31° corresponding to 101 Cr₂N. All Pole figures are plotted on a log intensity scale.

## Acknowledgments

The authors acknowledge funding from the Knut and Alice Wallenberg foundation through the Wallenberg Academy Fellows program (KAW-2020.0196), the Swedish Research Council (VR) under Project No. 2021-03826, 2025-03680 (PE), and the Swedish Foundation for Strategic Research (APR23-0017, LF). Ion-beam analysis was performed at Tandem laboratory, financed by the Swedish Research Council VR-RFI under contract number 2019_00191. ToF-ERDA was performed by Dr. Robert Frost.

## Corresponding Authors

Lewis J. Adams: Department of Chemistry – Ångström Laboratory, Uppsala University, SE-75120 Uppsala, Sweden. https://orcid.org/0009-0003-7497-2179; Email: lewis.adams@kemi.uu.se

## Authors

**Lewis J. Adams:** Department of Chemistry – Ångström Laboratory, Uppsala University, SE-75120 Uppsala, Sweden.

**Sara Baserga:** Department of Chemistry – Ångström Laboratory, Uppsala University, SE-75120 Uppsala, Sweden.

**Laurent Souqui:** Department of Chemistry – Ångström Laboratory, Uppsala University, SE-75120 Uppsala, Sweden.

**Enji Sadek:** Department of Chemistry – Ångström Laboratory, Uppsala University, SE-75120 Uppsala, Sweden.

**Linus von Fieandt:** Department of Chemistry – Ångström Laboratory, Uppsala University, SE-75120 Uppsala, Sweden. AB Sandvik Coromant, SE-12679 Hägersten, Sweden.

**Per Eklund:** Department of Chemistry – Ångström Laboratory, Uppsala University, SE-75120 Uppsala, Sweden.

## Author Contributions

CRediT: **Lewis J.Adams** conceptualization, data curation, formal analysis, investigation, supervision, visualization, writing- original draft, writing- review & editing; **Sara Baserga** data curation, formal analysis, investigation, visualization, writing- review & editing; **Laurent Souqui:** data curation, formal analysis, writing- review & editing; **Enji Sadek** data curation; **Linus von Fieandt:** data curation, formal analysis, visualization, resources, writing - review & editing; **Per Eklund** supervision, funding acquisition, resources, writing – review & editing.

## Notes

The authors declare no competing financial interest

## References

1. Aktarer SM, Sert Y, Küçükömeroğlu T. Investigation of structural, hardness, adhesion, and tribological properties of CrN and AlCrN coatings deposited on cylinder liner. Mater Des. 2025 May;253:113972. doi:10.1016/j.matdes.2025.113972

2. Hidalgo-Badillo JA, Hernández-Casco I, Herrera Hernández H, Soriano-Vargas O, Contla-Pacheco AD, González Morán CO, et al. A Tribological Study of CrN and TiBN Hard Coatings Deposited on

Cobalt Alloys Employed in the Food Industry. Coatings. 2024 Oct 7;14(10):1278. doi:10.3390/coatings14101278

3. Poorzal P, Elmkhah H, Mazaheri Y. Correlation between nanoindentation response and wear characteristics of CrN-based coatings deposited by an Arc-PVD method. Int J Appl Ceram Technol. 2022 May 12;ijac.14079. doi:10.1111/ijac.14079

4. Li K, Shao L, Li W, Shang L, Li R, Zhang Y, et al. Superior wear resistance of CrN film by PVD/HVOF structure design. Tribol Int. 2025 Sep;209:110753. doi:10.1016/j.triboint.2025.110753

5. He Y, Gao K, Yang H, Pang X, Volinsky AA. Nitrogen effects on structure, mechanical and thermal fracture properties of CrN films. Ceram Int. 2021 Nov;47(21):30729–40. doi:10.1016/j.ceramint.2021.07.252

6. Buchwald VF, Scott ERD. First Nitride (CrN) in Iron Meteorites. Nat Phys Sci. 1971 Oct;233(41):113–4. doi:10.1038/physci233113a0

7. Gharavi MA, Gambino D, le Febvrier A, Eriksson F, Armiento R, Alling B, *et al*. High thermoelectric power factor of pure and vanadium-alloyed chromium nitride thin films. Mater Today Commun. 2021 Sep;28:102493. doi:10.1016/j.mtcomm.2021.102493

8. Eklund P, Kerdsongpanya S, Alling B. Transition-Metal-Nitride-Based Thin Films as Novel Thermoelectric Materials. In: Mele P, Narducci D, Ohta M, Biswas K, Morante J, Saini S, et al., editors. Thermoelectric Thin Films [Internet]. Cham: Springer International Publishing; 2019 [cited 2026 Jan 15]. p. 121–38. Available from: http://link.springer.com/10.1007/978-3-030-20043-5_6 doi:10.1007/978-3-030-20043-5_6

9. Titirici M, Baird SG, Sparks TD, Yang SM, Brandt-Talbot A, Hosseinaei O, et al. The sustainable materials roadmap. J Phys Mater. 2022 Jul 1;5(3):032001. doi:10.1088/2515-7639/ac4ee5

10. Liu S, Qin Y, Wen Y, Shi H, Qin B, Hong T, et al. Efforts Toward the Fabrication of Thermoelectric Cooling Module Based on N-Type and P-Type PbTe Ingots. Adv Funct Mater. 2024 Jun;34(26):2315707. doi:10.1002/adfm.202315707

11. Warcholinski B, Gilewicz A, Kuprin AS, Kolodiy IV. Structure and properties of CrN coatings formed using cathodic arc evaporation in stationary system. Trans Nonferrous Met Soc China. 2019 Apr;29(4):799–810. doi:10.1016/S1003-6326(19)64990-3

12. Semenchuk NV, Kolubaev AV, Sizova OV, Denisova YuA, Leonov AA. Structure and Properties of Multilayer CrN/TiN Coatings Obtained by Vacuum-Arc Plasma-Assisted Deposition on Copper and Beryllium Bronze. Russ Phys J. 2023 Dec;66(10):1077–86. doi:10.1007/s11182-023-03045-5

13. Leonov AA, Denisova YuA, Denisov VV, Savostikov VM, Syrtanov MS, Pirozhkov AV, et al. Physical and Mechanical Properties of CrN/AlN Coating Obtained by Vacuum-Arc Deposition with Alternative Separation of Hard Substance Flows. Russ Phys J. 2024 Jan;66(11):1152–7. doi:10.1007/s11182-023-03056-2

14. Singh NK, Hjort V, Honnali SK, Gambino D, le Febvrier A, Ramanath G, et al. Effects of W alloying on the electronic structure, phase stability, and thermoelectric power factor in epitaxial CrN thin films. J Appl Phys. 2024 Oct 21;136(15):155301. doi:10.1063/5.0226046

15. Gharavi MA, Kerdsongpanya S, Schmidt S, Eriksson F, Nong NV, Lu J, et al. Microstructure and thermoelectric properties of CrN and CrN/$Cr_2$N thin films. J Phys Appl Phys. 2018 Sep 5;51(35):355302. doi:10.1088/1361-6463/aad2ef

16. Liu YC, Hsiao SN, Chen YH, Hsieh PY, He JL. High-Power Impulse Magnetron Sputter-Deposited Chromium-Based Coatings for Corrosion Protection. Coatings. 2023 Dec 18;13(12):2101. doi:10.3390/coatings13122101

17. Dasgupta A, Kuppusami P, Lawrence F, Raghunathan VS, Antony Premkumar P, Nagaraja KS. Plasma assisted metal-organic chemical vapor deposition of hard chromium nitride thin film coatings using chromium(III) acetylacetonate as the precursor. Mater Sci Eng A. 2004 Jun;374(1–2):362–8. doi:10.1016/j.msea.2004.03.021

18. Schuster F, Maury F, Nowak JF, Bernard C. Characterization of chromium nitride and carbonitride coatings deposited at low temperature by organometallic chemical vapour deposition. Surf Coat Technol. 1991 Sep;46(3):275–88. doi:10.1016/0257-8972(91)90170-2

19. Hampden-Smith MJ, Kodas TT, editors. The chemistry of metal CVD. Weinheim New York: VCH; 1994. 1 p. doi:10.1002/9783527615858

20. Lang H, Dietrich S. Metals – Gas-Phase Deposition and Applications. In: Comprehensive Inorganic Chemistry II [Internet]. Elsevier; 2013 [cited 2026 Feb 22]. p. 211–69. Available from: https://linkinghub.elsevier.com/retrieve/pii/B9780080977744004125 doi:10.1016/B978-0-08-097774-4.00412-5

21. Ogden JS, Wyatt RS. Matrix isolation and mass spectrometric studies on the vaporisation of chromium(III) chloride. The characterisation of molecular CrCl4 and CrCl3. J Chem Soc Dalton Trans. 1987;(4):859. doi:10.1039/dt9870000859

22. Mannei E, Asedegbega-Nieto E, Ayari F. Thermal treatment of anhydrous chromium (III) chloride: Thermodynamic study. Thermochim Acta. 2022 Jan;707:179102. doi:10.1016/j.tca.2021.179102

23. Wakefield GF. Chromium Coatings Prepared by Chemical Vapor Deposition. J Electrochem Soc. 1969;116(1):5. doi:10.1149/1.2411774

24. Mazille HMJ. Chemical vapour deposition of chromium onto nickel. Thin Solid Films. 1980 Jan;65(1):67–74. doi:10.1016/0040-6090(80)90058-9

25. Abelson JR, Girolami GS. New strategies for conformal, superconformal, and ultrasmooth films by low temperature chemical vapor deposition. J Vac Sci Technol A. 2020 May 1;38(3):030802. doi:10.1116/6.0000035

26. Hulkko JG. Muspel and Surtr: CVD system and control program for WF6 chemistry [Internet]. Available from: http://www.diva-portal.org/smash/record.jsf?pid=diva2%3A1317466&swid=-3177

27. Jin Q, Wang Z, Zhang Q, Zhao J, Cheng H, Lin S, et al. Structural twinning-induced insulating phase in CrN (111) films. Phys Rev Mater. 2021 Feb 22;5(2):023604. doi:10.1103/PhysRevMaterials.5.023604

28. Le Febvrier A, Tureson N, Stilkerich N, Greczynski G, Eklund P. Effect of impurities on morphology, growth mode, and thermoelectric properties of (1 1 1) and (0 0 1) epitaxial-like ScN films. J Phys Appl Phys. 2019 Jan 16;52(3):035302. doi:10.1088/1361-6463/aaeb1b

29. Yamauchi R, Hamasaki Y, Shibuya T, Saito A, Tsuchimine N, Koyama K, et al. Layer matching epitaxy of NiO thin films on atomically stepped sapphire (0001) substrates. Sci Rep. 2015 Sep 24;5(1):14385. doi:10.1038/srep14385

30. le Febvrier A, Gambino D, Giovannelli F, Bakhit B, Hurand S, Abadias G, et al. p -type behavior of CrN thin films via control of point defects. Phys Rev B. 2022 Mar 16;105(10):104108. doi:10.1103/PhysRevB.105.104108

31. Balasubramanian K, Khare SV, Gall D. Energetics of point defects in rocksalt structure transition metal nitrides: Thermodynamic reasons for deviations from stoichiometry. Acta Mater. 2018 Oct;159:77–88. doi:10.1016/j.actamat.2018.07.074

32. Pankratova D, Yusupov K, Vomiero A, Honnali SK, Boyd R, Fournier D, et al. Enhanced Thermoelectric Properties by Embedding Fe Nanoparticles into CrN Films for Energy Harvesting Applications. ACS Appl Nano Mater. 2024 Feb 9;7(3):3428–35. doi:10.1021/acsanm.3c06054

33. Tsai MH, Sun SC, Chiu HT, Tsai CE, Chuang SH. Metalorganic chemical vapor deposition of tantalum nitride by tertbutylimidotris(diethylamido)tantalum for advanced metallization. Appl Phys Lett. 1995 Aug 21;67(8):1128–30. doi:10.1063/1.114983

34. Imhoff L, Bouteville A, De Baynast H, Remy JC. Evaluation and localization of oxygen in thin TiN layers obtained by RTLPCVD from TiCl4–NH3–H2. Solid-State Electron. 1999 Jun;43(6):1025–9. doi:10.1016/S0038-1101(99)00019-2

35. Lippitz A, Hübert Th. XPS investigations of chromium nitride thin films. Surf Coat Technol. 2005 Oct;200(1–4):250–3. doi:10.1016/j.surfcoat.2005.02.091

36. Réti F, Bertóti I, Mink G, Székely T. Surface reactions of chlorine with /gg-alumina. React Solids. 1987 Aug;3(4):329–36. doi:10.1016/0168-7336(87)80071-2

37. Fu W, Chai J, Kawai H, Maddumapatabandi T, Bussolotti F, Huang D, et al. Evidence of air-induced surface transformation of atomic step-engineered sapphire in relation to epitaxial growth of 2D semiconductors. Nat Commun. 2025 Sep 26;16(1):8488. doi:10.1038/s41467-025-63452-9

38. Mozafari E, Alling B, Steneteg P, Abrikosov IA. Role of N defects in paramagnetic CrN at finite temperatures from first principles. Phys Rev B. 2015 Mar 2;91(9):094101. doi:10.1103/PhysRevB.91.094101

39. le Febvrier A, Tureson N, Stilkerich N, Greczynski G, Eklund P. Effect of impurities on morphology, growth mode, and thermoelectric properties of (1 1 1) and (0 0 1) epitaxial-like ScN films. J Phys Appl Phys. 2019 Jan 16;52(3):035302. doi:10.1088/1361-6463/aaeb1b

40. Langford RM, Clinton C. In situ lift-out using a FIB-SEM system. Micron. 2004 Oct;35(7):607–11. doi:10.1016/j.micron.2004.03.002

41. Ström P, Primetzhofer D. Ion beam tools for nondestructive in-situ and in-operando composition analysis and modification of materials at the Tandem Laboratory in Uppsala. J Instrum. 2022 Apr 1;17(04):P04011. doi:10.1088/1748-0221/17/04/P04011

42. Arstila K, Julin J, Laitinen MI, Aalto J, Konu T, Kärkkäinen S, et al. Potku – New analysis software for heavy ion elastic recoil detection analysis. Nucl Instrum Methods Phys Res Sect B Beam Interact Mater At. 2014 Jul;331:34–41. doi:10.1016/j.nimb.2014.02.016

43. Xin B, le Febvrier A, Wang L, Solin N, Paul B, Eklund P. Growth and optical properties of CaxCoO2 thin films. Mater Des. 2021 Nov;210:110033. doi:10.1016/j.matdes.2021.110033